\documentclass[referee]{raa}
\usepackage{graphicx,times}
\usepackage{natbib}
\usepackage{amssymb,amsmath}
\bibpunct{(}{)}{;}{a}{}{,}

\usepackage[a4paper=true,dvipdfm=true,pagebackref=true]{hyperref}
\hypersetup{pdftitle = The title of my PDF, pdfauthor = My name, pdfsubject= The subject, pdfkeywords = keyword1 keyword2 keyword3}
\hypersetup{colorlinks = true, linkcolor = green, anchorcolor = red, citecolor = blue, filecolor = red, pagecolor = red, urlcolor = red}

\begin{document}

   \title{Diffuse emission of TeV Neutrinos and Gamma-rays from young pulsars by Photo-meson
   interaction in the galaxy }

 \volnopage{ {\bf 2013} Vol.\ {\bf X} No. {\bf XX}, 000--000}
   \setcounter{page}{1}

   \author{Zhi-Xiong Li\inst{1}, Gui-Fang
   Lin\inst{2,3},Wei-Wei Na\inst{4}}

   \institute{ Department of Physics, Yunnan University,
    Kunming 650091, China; {\it zhixiongli456789@gmail.com}\\
        \and
             Yunnan Astronomical Observatory, National Astronomical Observatories, Chinese Academy of Sciences,
             Kunming 650011, China\\
             \and
             Laboratory for the Structure and Evolution of Celestial
             Objects,Chinese Academy of Sciences,Kunming
             650011,China\\
             \and
              Department of Physics, Yuxi Normal University,
              Yuxi 653100, China;{\it naww@yxnu.net}\\
\vs \no
   {\small Received 2013 ; accepted 2013 }
}

\abstract{It's generally believed that young and rapidly rotating
pulsars are important sites of particle's acceleration, in which
protons can be accelerated to relativistic energy above the polar
cap region if the magnetic moment is antiparallel to the spin
axis($\vec{\mu}\cdot\vec{\Omega}<0$).
 To obtain the galactic diffusive neutrinos and gamma-rays for TeV, firstly,we use Monte Carlo(MC) method to generate a sample of young pulsars
 with ages less than $10^6$ yrs in our galaxy ; secondly, the neutrinos and high-energy gamma-rays can be produced through photomeson process with
  the interaction of energetic protons and soft X-ray photons  ($p+\gamma\rightarrow \Delta^+\rightarrow
n+\pi^+/p+\pi^0$) for single pulsar, and these X-ray photons come
from the neutron star surface.
   The results suggest that the diffusive TeV flux of neutrinos are  lower than background flux, which indicated it is difficult to be detected by the current neutrino telescopes.
--- stars: neutron
--- neutrinos and gamma-rays: stars --- elementary: neutrinos}

   \authorrunning{Z.-X. Li et al. }            
   \titlerunning{Diffuse emission of Neutrinos and TeV Gamma-rays from young pulsars}  
   \maketitle

%
\section{Introduction}           
\label{sect:intro} Very high-energy[VHE,$E>100GeV$] neutrinos and
gamma-rays from astrophysical objects can provide a clear indication
of the origins of galactic and extragalactic cosmic rays.
 The radiations probably come from gamma-ray
 bursts,active galactic nuclei,pulsars etc.
 In our Galaxy most  high-energy gamma-rays and neutrinos are associated with
 pulsars, supernova remnant and pulsar nebula(\citealt{Bednarek+etal+2005};\citealt{Kistler+etal+2007}).
 Young pulsars are persistent and periodic sources; they also have shorter distances to Earth than extragalactic sources. If they emit high-energy neutrinos/gamma-rays, we can spend longer time
 searching for the signals.If the
TeV neutrinos and gamma-rays can be detected from pulsars, it will
help improve our knowledge
 about the hadronic process taking place in the
 magnetosphere.

  Recently, some studies (\citealt{Link+Burgio+2005,Link+Burgio+2006};\citealt{Bhadra+Dey+2009};\citealt{Jiang+etal+2007})
 have  shown that if pulsar(age
less than  $10^6yr$) magnetic moment is antiparallel to the spin
axis ($\vec{\mu}\cdot\vec{\Omega}<0$),$\Delta$-resonance will be
produced when pulsars-accelerated ions interact with these thermal
radiation field. Then high-energy neutrinos and gamma-rays decay by
$\Delta$-resonance. Link\&Burgio(hereafter LB) considered that near
the surface of the neutron stars, the protons or heavier ions can be
accelerated by polar caps to PeV energy.When the thermal radiation
flied of pulsar interact with accelerated ions with PeV energy,
$\Delta$-resonance state may occur.
 This process is effective,and pions subsequently decay to muon neutrinos or gamma-rays .
 LB calculated  energy spectrum of muon neutrinos from some pulsars and estimated the
 event rates at Earth.
 The spectrum sharp rise at $\sim$50 TeV, corresponding to the
 onset of the resonance. The flux drops with neutrino
 energy as $\epsilon_{\nu}^{-2}$up to an upper energy cut-off that is
 determined by either kinematics or the maximum energy to which
 protons are accelerated. Bhadra \& Dey(2009)estimated TeV gamma-ray flux
 at the Earth from a few nearby young pulsars and compared with the
observation, they found that proper consideration of the effect of
the polar cap geometry in flux calculation is important,then they
revised the event rates.

  In this paper we find the $\eta$ (ratio of polar cap area to neutron star surface area)is
  double the
  Bhadra \& Dey give.We use revised polar cap geometry to
calculate TeV neutrinos and  gamma-rays spectrum from some young
pulsars,and estimate the
 event rates at Earth.  In order to estimate diffuse
TeV neutrinos and gamma-rays radiation from Galactic Plane, a galaxy
pulsar sample is required. It is the better way to simulate a pulsar
sample with MC method
(\citealt{Cheng+etal+1998};\citealt{Zhang+etal+2000};\citealt{Jiang+etal+2007}).
 At last,we obtain the TeV  diffuse  Neutrinos and Gamma-rays flux from young pulsars in our
 galaxy.

  In section 2,we reviewed the model of LB. In section
  3,we calculate neutrino and gamma-ray spectrum from single pulsar by revised polar cap
  geometry.In section 4 and 5,we obtain a sample of
young pulsars in our galaxy with Monte Carlos method,
 then estimate the diffuse neutrinos and gamma-rays emission from Galactic young pulsars.At last the discussion and
conclusion are presented.

\section{The model of neutrinos and gamma-rays from young pulsars}
\label{sect:Tev}

Neutron stars have enormous magnetic fields($B\geq10^{12}G$) and
high rotation rates(tens of Hertz)which act as a very powerful
generator. Charges will be stripped off the highly conductive
surface and accelerated somewhere above the stellar surface. The
acceleration of particles mechanisms from pulsars generally has been
divided into the  polar gap model
(\citealt{Ruderman+etal+1975};\citealt{Arons+etal+1979};\citealt{Daugherty+etal+1996};\citealt{Gonthier+etal+2002})
and outer-gap
model(\citealt{Zhang+etal+2004};\citealt{Cheng+etal+1986}). In the
former, particles are accelerated in charge-depleted zone region
near the magnetic pole of the neutron star. In the outer-gap, it
will take place in the vacuum gaps between the
 neutral line and the last open line in the
 magnetosphere. Therefore, acceleration region in the polar-gap model is close to the neutron star surface,whereas in the outer-gap model it is near to the light cylinder.

In polar gap  model, because of existing large rotation-induced
electric fields, particles can be extracted from the polar cap
surface,
 and then be accelerated, finally form the primary beam. The
 potential drop cross the field of a pulsar from the magnetic
 pole to the last field line open to infinity is
 $\Delta\phi=B_{s}R^{3}\Omega^{2}/2c^{2}$(\citealt{Goldreich+etal+1969}),where $B_{s}$ is the
 strength of the dipole component of the field at the magnetic poles ($B_{s}\sim10^{12}G$)
 and  $R=10^6R_{6}$ is the radius of the neutron star and $p_{m}$ is the spin period in milliseconds ,$\Omega=2\pi/p$ is the angular
 velocity(where $p$ is the period ) and $c$ is the  speed of the
 light. The magnitude of the potential drop  $\Delta\phi$ could be
 $7\times10^{18}B_{12}P_{ms}^{-2}$ volts ($B_{s}\equiv
 B_{12}\times10^{12}$)(\citealt{Goldreich+etal+1969}).L B consider that If electric filed is no or little
 screening and  $\mu\cdot\Omega<0$ ( expected to hold for half of total
 pulsars), ions and protons will be accelerated to PeV energy near the surface of
 pulsars.$\mu$ is magnetic moment, $\Omega$ is angular velocity.
  Protons or ions  accelerated by pulsars will interact with the thermal radiation of pulsars. If
  they have sufficient energy and exceed the threshold energy for the $\Delta$-resonance state($\Delta^{+}$ is an excited state of proton,with
 a mass of 1232MeV).The $\Delta$-resonance state may occurr. The threshold condition
 for the production of $\Delta$-resonance state in $p-\gamma$
 interaction is given by
 \begin{equation}
 \epsilon_{p}\epsilon_{\gamma}(1-cos\theta_{p\gamma})\geq 0.3
 GeV^{2},
 \end{equation}
 where $\epsilon_{p}$ and $\epsilon_{\gamma}$ are the proton and photon
 energy, and $\theta_{p\gamma}$ is the incidence angle between the
 proton and the photon in the lab frame.Young pulsars have typically
  temperatures of $T_{\infty}\simeq 0.1 keV$, and photo energy
  $\epsilon_{\gamma}=2.8kT_{\infty}(1+Z_{g})\sim 0.4 keV$, where
  $Z_{g}\approx 0.4$ is the gravitational red shift and
  $T_{\infty}$is the surface temperature measured at infinity.
In a young pulsar's atmosphere, the condition of production
$\Delta$-resonance is $B_{12}P_{ms}^{-2}T_{0.1keV}\geq
3\times10^{-4}$(\citealt{Link+Burgio+2005};\citealt{Link+Burgio+2006}),
where $T_{0.1keV}\equiv
 (kT_{\infty}/0.1keV)$,$T_{\infty}\sim0.1keV$ is typical surface
 temperature of young pulsars. This condition holds for many young pulsars,so $\Delta$-resonance could
 existed in many pulsar's atmosphere.
Gamma-rays and neutrinos subsequently decayed through following
channels

\[p + \gamma \rightarrow \Delta^{+} \rightarrow \left\{\begin{array}{ll}
p + \pi^{o} \rightarrow p + 2\gamma \\
n\pi^{+} \rightarrow n + e^{+} + \nu_{e} + \nu_{\mu} +
\bar{\nu_{\mu}}.
\end{array}
\right . \]

 In this part, based on the L B model;we
 estimated emitting neutrinos and gamma-rays from pulsars. The flux of protons accelerated
  by polar gap  can be estimated
  \begin{equation}
 I_{pc}=c f_{d}(1-f_{d}) n_{o} A_{pc},
\end{equation}
where $n_{o}(r)\equiv B_{s}R^{3} \Omega /(4 \pi Ze c r^{3})$ is the
Goldreich-Julian density of ions in distance $r$,$f_{d}<1$ is the
fraction depletion in the space charge in the acceleration region.
It's a model dependent quantity ($f_{d}=0 $ correspond to no
depletion and $f_{d}=1 $ is full depletion), so the density in the
depleted gap can be written as $f_{d}(1-f_{d}) n_{o}$,where
$A_{pc}=\eta 2 \pi R^{2}$ is the polar cap area, $\eta$ is the ratio
of polar cap area to the half of neutron star surface area, when
$\eta=1$ is the half of neutron star surface area. In LB(2006)and
Jiang(2007)they use hemisphere surface area to calculate.The typical
radius $r_{pc}= R (\Omega R/c)^{1/2}$
(\citealt{beskin+etal+1993}),the polar gap surface is $A_{pc}=\pi
r_{pc}^2=\pi\Omega R^3/c$,so $\eta=\Omega R/(2c)$.  So the maybe
$\eta=\Omega R/(2c)$ is more appropriate. For young pulsars with
surface temperature $T_{\infty}$, the photon density close to the
neutron star surface is
$n_{\gamma}(R)=(a/2.8k)[(1+z_{g})T_{\infty}]$, $a$ is the
Stefan-Boltzman constant. Numerically
$n_{\gamma}(R)\simeq9\times10^{19}T_{0.1kev}^3$. At radial distance
$r$, photon density will be $n_{\gamma}(r)=n_{\gamma}(R)(R/r)^2$.
The probability that a PeV energy proton staring from the pulsar
surface will produce $\Delta$ particle by interacting thermal filed
which is given by(link\&Burgio 2005) $p_{c}=1-\int_{R}^{r}p(r)$.
Where $dP/P=-n_{\gamma}(r)\sigma_{p\gamma}dr$. Thus the total flux
of gamma ray/neutri generated in pulsars from the $\Delta^{+}$
resonance is
\begin{equation}
I =  2 c \xi A_{pc} f_{d} (1-f_{d}) n_{o} P_{c},
\end{equation}
where $\xi$ is $4/3$ and $2/3$ for gamma rays and muon
neutrinos,respectively. At the distance of $d$, the phase averaged
gamma ray/neutrino flux at Earth from a pulsar is
\begin{equation}
\phi \simeq  2 c \xi \zeta \eta f_{b} f_{d} (1-f_{d})
n_{o}\left(\frac{R}{d}\right)^{2}P_{c},
\end{equation}
 where $f_{b}$ is duty cycle of neutrino or gamma-ray beam, $\zeta$ is the effect due to neutrino oscillation (the
decays of pions and their muon daughters result in initial flavor
ratios $\phi_{\nu_{e}}:\phi_{\nu_{\mu}}:\phi_{\nu_{\tau}}$ of nearly
$1:2:0$ but at large distance from the source the flavor ratios is
expected to become $1:1:1$ due to maximal mixing of $\nu_{\mu}$ and
$\nu_{\tau}$). $\zeta =1$ and 1/2 is gamma rays and muon neutrinos
respectively. We want to obtain the spectrum of neutrinos and
gamma-rays,  using the following differential from
\begin{equation}
\frac{d\phi_{\nu}}{d\epsilon_{\nu}}= 2 c \xi \zeta \eta f_{b} f_{d}
(1-f_{d})
n_{o}\left(\frac{R}{d}\right)^{2}\frac{dP_{c}}{d\epsilon_{\nu}},
\end{equation}
Taking $f_{d}=1/2$ and $Z=A=1$ for estimating upper limits on the
flux. The neutrino and Gamma-rays energy flux are estimated as
\begin{equation}
\frac{d\phi_{\nu}}{d\epsilon_{\nu}}=3\times10^{-8}\xi \zeta \eta
f_{b}B_{12}p^{-1}_{ms}d^{-2}_{kpc}T_{0.1kev}\frac{dP_{c}}{dx}.
\end{equation}
Where $d_{kpc}$ is the distance of pulsar to earth, and $dP_{c}/dx$
is the probability of proton converting to $\Delta^{+}$ per unit
energy interval; the details have been shown in equation(26) at Link
\& Burgio(2006).

\section{ TeV Neutrinos and Gamma-rays  spectrum}
 \label{sect:data}
 ~~~~We use the equation (6) to estimate flux of neutrinos and gamma-rays
and take Z=1 and $f_{d}=1/2$ throughout this work. Taking linear
$(\gamma=1)$ and quadratic $(\gamma=2)$ proton acceleration laws.
Linear acceleration is corresponding to an accelerating filed which
is constant space and quadratic accelerating field  grows linearly
with height above the star. We calculate the neutrinos and
gamma-rays energy flux of crab,vela,PSR B1509-58 and PSR
B1706-44,when $\eta=\Omega R/2c$ showed in the fig.1 and fig.2. The
parameter of source presents in Table.1.
\begin{figure}
   \centering
   \includegraphics[width=14.0cm, angle=0]{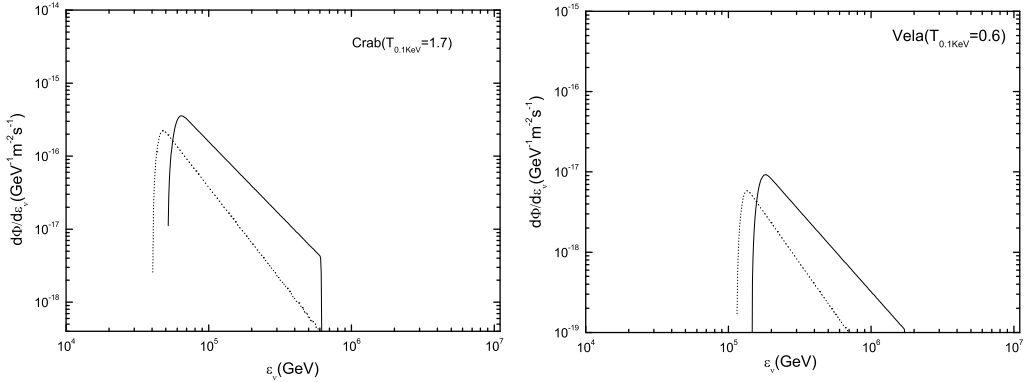}
  \caption{ The neutrino flux displayed for crab,vela, the case of linear(solid line
 ) and quadratic proton acceleration (doted line).}
   \label{Fig1}
   \end{figure}

\begin{figure}
   \centering
   \includegraphics[width=14.0cm, angle=0]{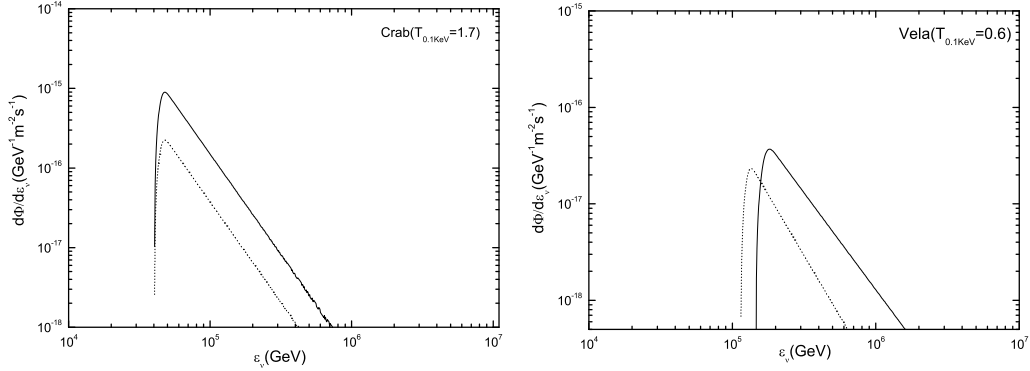}
   \caption{The gamma-ray flux displayed for crab,vela,the case of linear(solid line) and quadratic proton acceleration (doted line). }
   \label{Fig2}
   \end{figure}

\begin{table}
\bc
\begin{minipage}[]{100mm}

\caption[]{ The parameter of some pulsars}\end{minipage}

\small
 \begin{tabular}{ccccccccccc}
  \hline\noalign{\smallskip}
Source         & {\em d}         &{\em p}      & {\em $B_{12}$}    & {\em $T_{0.1 keV}$}   & $f_{b}$   & $\frac{dN}{dAdt}$  (LB06)    & $\frac{dN}{dAdt}$\\
               &    kpc          &    ms       &     G              &                      &           &   $km^{-2}yr^{-1}$           & $km^{-2}yr^{-1}$     \\
  \hline\noalign{\smallskip}
 Crab    &   2                   &   33      &      3.8                  & $\sim 1.7$       &   0.14   &   45                         &   0.151   \\
     Vela     &  0.29                 &   89      &      3.4                  &   0.6       &   0.04   &   25                         &   0.0313      \\
$B 1706-44$   &  1.8                  & 102       &       3.1                 & 1           &   0.13   &   1                          &   0.0038     \\
$B 1509-58$   &  4.4                  & 151       &      0.26                 & 1           &   0.26   &   5                          &   0.0057   \\
  \noalign{\smallskip}\hline
\end{tabular}
\ec
\end{table}
\begin{table}
\bc
\begin{minipage}[]{100mm}

\caption[]{The integral Tev gamma-ray fluxes comparison between the
predicated and the observed.The observed upper limits for Crab,
Vela,PSR B1509-58 and PSR B 1706-44
(\citealt{Aharonian+etal+2006a})}
\end{minipage}

\small
 \begin{tabular}{ccccccccccc}
  \hline\noalign{\smallskip}
source          & $\eta=1$                      & $\eta=\Omega R /(2c)$        & observed upper limit of integral flux   \\
                  & $10^{-15}$$cm^{-2}$$s^{-1}$   & $10^{-15}$$cm^{-2}$$s^{-1}$  & $10^{-15}$$cm^{-2}$$s^{-1}$  \\
  \hline\noalign{\smallskip}
     Crab      &  1038       &  3.297  & 8(56)  \\
     Vela      &   204.98    &  0.242   & 10(20) \\
$B 1706-44$    &   67.4773   &  0.0692  &10(20)  \\
$B 1509-58$    &  69.675     &  0.0484  & 10(20)\\
\noalign{\smallskip}\hline
\end{tabular}
\ec
\end{table}

 For either acceleration law, the spectrum turns sharply at
$\varepsilon_{\nu}\simeq 70_{0.1keV}^{-1}TeV$ corresponding to the
onset of $\Delta$-resonance. At the top of energy it drops
approximately as $\epsilon_{\nu}^{-2}$,as the phase space because
the conversion becomes restricted. The  flux is lower about 3
magnitude when we use pulsar surface to calculate .But it is more
consistent with the  observed upper limits of gamma-ray
fluxes.numerical values of the integral TeV gamma-ray fluxes are
obtained for pulsars listed in Table 2.

In table 2, The gamma-rays flux estimated with $\eta=1$is obviously
higher than the observed upper limits. But with $\eta=\Omega R/2c$
it is a little lower and more consistent with the observed upper
limits; Thus taking $\eta=\Omega R /(2c)$ is more reasonable.

Large-area neutrino detectors use the Earth or ice as a medium for
conversion of muon neutrino to muon, detecting the Cerenkov light
through the high energy upward-moving  muons produced by neutrino
interactions below a detector on the surface of earth. We can use
the flux of neutrino estimate the count rate in the detector. The
conversion probability in the Earth is(\citealt{Gaisser+etal+1995}).
\begin{equation}
P_{\nu_{\mu}\rightarrow\mu}=1.3\times10^{-6}(\epsilon_{\nu}/1 TeV),
\end{equation}
The muon event rate is
\begin{equation}
\frac{dN}{dAdt}=\int
d\epsilon_{\nu}\frac{d\phi_{\nu}}{d\epsilon_{\nu}}P_{\nu_{\mu}\rightarrow\mu}.
\end{equation}
In Table 1,expected count rates what we estimated present in the
last column of for the crab,vela,PSR B1509-58 and PSR B1706-44,use
$L=0.1$ and linear acceleration. Expected count rates (LB06)are
shown in the penultimate column.Compared with the expected count
rates shown in the last column, expected count rates(LB06)are
higher. It means that the detectable by IceCube will not look
bright.(\citealt{Abbasi+etal+2012}) It shows that IceCube data
severely constrain these optimistic predictions of LB, and
(\citealt{Bhadra+Dey+2009}) pulsars are unlikely to be strong source
of TeV neutrino. So proper consideration of the effect of polar cap
geometry in flux calculation is important.

\section{The Diffuse Neutrinos and Gamma-rays emission from Galactic young pulsars}
   Over 1800 radio pulsars are known (Manchester et al 2005), more than 400 pulsars ages are less than $10^6yr$,
most of which are candidates. If the magnetic moment is antiparallel
to the rotating axis($\mu\cdot\Omega<0$),ions can be accelerated in
the charge-depleted gap near the star surface.So young pulsars are
also potential neutrino sources, although some of them are likely
weak source, the total contributions maybe significant. Therefore we
use MC method simulate pulsars sample with ages less than
$10^6yr$,then determine which is neutrino pulsars in sample. We also
estimate the diffuse neutrino flux and gamma-ray flux. In other
words,we will cumulate all potential neutrino pulsars in the sample.

\subsection{Pulsars \textbf{Sample simulated by Monte Carlo method}}
We  produce the Galactic young pulsar population by following
assumptions.(e.g.\citealt{Sturner+etal+1996};\citealt{Cheng+etal+1998};\citealt{Zhang+etal+2000};\citealt{Zhang+etal+2004};
\citealt{Jiang+etal+2006}).
 Pulsar sample is simulated  by MC method which is described by following steps in this paper.(see \citealt{Jiang+etal+2007} for
 detail).

\begin{figure}
   \centering
   \includegraphics[width=14.0cm, angle=0]{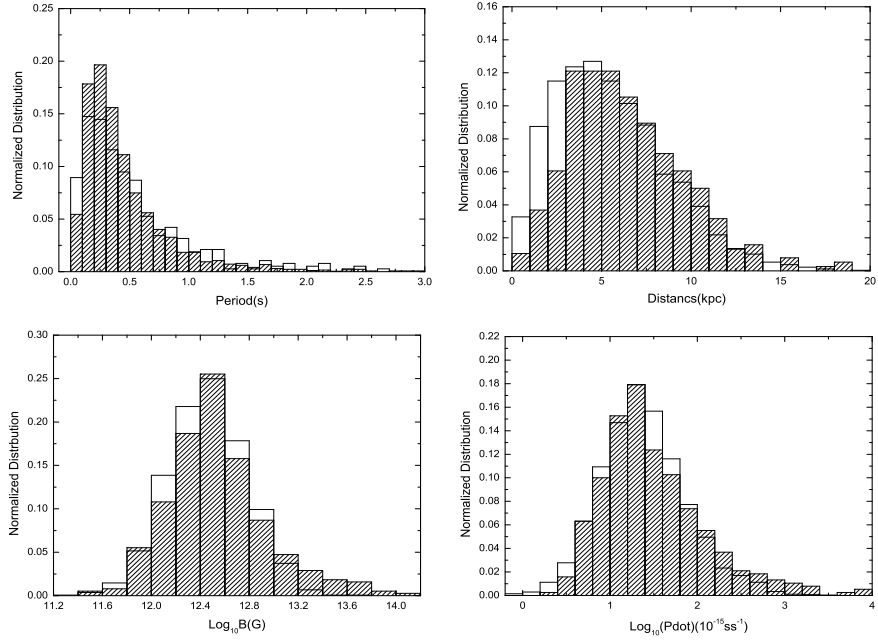}
  \caption{Normalized histogram distribution of pulsar period, distance,surface
magnetic field and period derivation. Shadow histogram represent
observations sample sample. Solid histogram represent simulated
sample.  }
   \label{Fig3}
   \end{figure}

\begin{figure}
   \centering
   \includegraphics[width=14.0cm, angle=0]{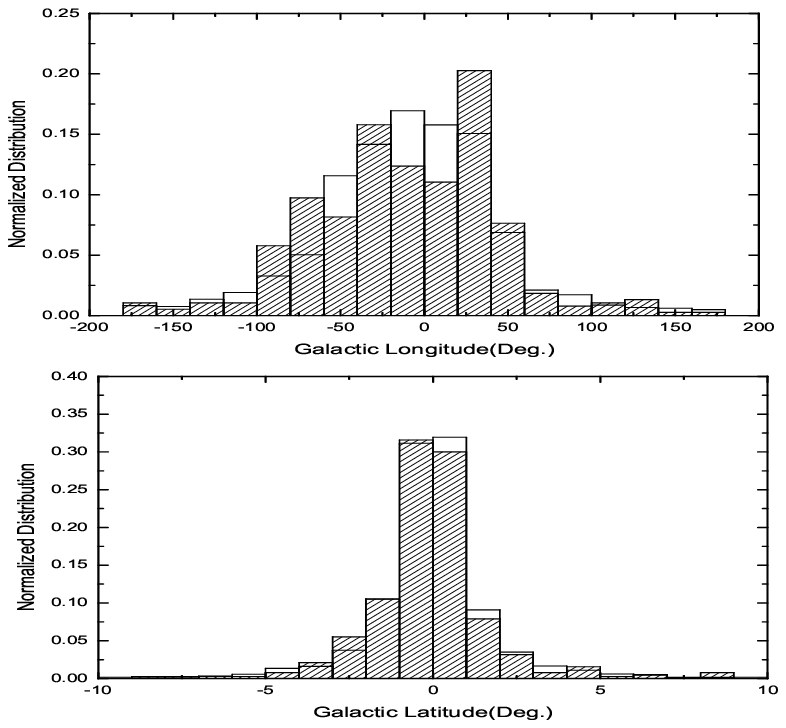}
  \caption{Normalized histogram
distribution of pulsars in Galactic Longitude and Galactic Latitude.
Shadow histogram represents observations sample. Solid histogram
represents simulated sample.}
   \label{Fig4}
   \end{figure}

Following the steps, we use the conventional assumption for Galactic
pulsars birth rate ($\sim 2/100 yr $) and enlarge 10 times, generate
about 160000 pulsars in Galactic plane during past millions years.
In this sample about 3890 pulsar could be detected radio.A lot of
pulsars are not be observed because of selection effect. Compared
with the ATNF catalogue pulsars, our simulated sample is consistent
with the distribution of the observation. The normalized histogram
distribution of pulsar period, pulsar surface magnetic field and
distance are shown in fig.3. The shadow histogram represents the
ANTF sample,the solid histogram represents simulated sample. We also
compare the distribution Galactic Longitude and Galactic Latitude
with the observed sample which is shown in fig.4. The shadow
histogram represents the ANTF sample,solid histogram represents the
simulated sample. Obviously,most pulsars distribute in Galactic
Latitude $|b|<5^{\circ}$ region. So our simulated is succeed.

\subsection{The Diffuse Emission of Neutrinos and Gamma-ray from
Young Pulsars}
 The diffuse neutrinos and gamma-rays emission, both in Galactic and extragalactic are
 very interest for astrophysics, particle physics, and cosmology. The
diffuse Galactic emission is produced by interactions of cosmic
rays, mainly protons or electrons interact with the interstellar gas
(via $\pi^0$-production and bremsstrahlung) and radiation field (via
inverse compton scattering). We estimate the
 gamma-rays and neutrinos flux$(d\phi_{v,i})/d\epsilon_\nu$ by using
 equation(6). Then we can obtain the flux of neutrinos and
 gamma-rays from all pulsars

 \begin{equation}
\phi(\epsilon_\nu)=\sum^{N}_{i=1}\frac{d\phi_{v,i}}{d\epsilon_{\nu}d\Omega_i
},
\end{equation}

where N is the number of gamma-ray or neutrinos pulsars, and
$d\Omega_i$ is the solid angle for the i pulsar, $d\Omega_i=4\pi
f_{b,i}\sim1$. The results of neutrino energy flux are shown in
fig.5. The energy flux sharply increase at about
$50TeV$,corresponding to the onset of the resonance. Most energy is
emitted between 50TeV and 0.8TeV.
 After the onset of the resonance,the spectrum drops approximately as $\epsilon_{\nu}^{-2}$, because the
phase space for conversion becomes restricted. Just like the
spectrum for a single pulsar,it is a typical of first-order Fermi
acceleration(\citealt{Abbasi+etal+2012}).  The linear proton
acceleration is about 5 times than quadratic acceleration, so the
linear proton acceleration is the main acceleration. From the figure
5,it will be not easier to search for neutrino signals, because it
only exceeds the lower limit atmospheric neutrino flux's background
a little. For comparison, we also give the predicted fluxes with
sensitivities of AMANDA-II, ANTARES, and IceCube detectors. AMANDA
detector reached a sensitivity of $8.9\times10^{-8}GeV
cm^{-2}s^{-1}sr^{-1}$ after the first 4 years of operation in
2000-2003 (\citealt{Hill+2006}). The ANTARES detector sensitivity
reached $7.8\pm0.99\times10^{-8}GeV cm^{-2}s^{-1}sr^{-1}$ after 1
year of data on diffuse flux, and $3.9\pm0.7\times10^{-8}GeV
cm^{-2}s^{-1}sr^{-1}$ after 3 years of data on diffuse flux
(\citealt{Montaruli+2005}).The IceCube detectors have reached a
sensitivity of $(2-7)\times10^{-9}GeV cm^{-2}s^{-1}sr^{-1}$ after 3
years of operation (\citealt{Ribordy+etal+2006}). It sees clearly
that it not easier to detect the signals of neutrino flux from young
pulsars, because the neutrino flux energy is lower than the
background atmospheric neutrino flux. From Figure 5, we also compare
the predicted diffuse neutrino fluxes from Active Galactic Nuclei
(AGNs)(\citealt{Stecker+etal+2005})and Gamma Ray Bursts
(GRBs)(\citealt{Liu+Wang+2013}). Whereas we can find the diffuse
neutrino fluxes from AGNs are larger than young pulsars and GRBs at
the energy from 60TeV to 3PeV. The diffuse neutrino fluxes from GRBs
and young pulsars at the ranges from 100TeV to 1PeV almost the same
and lower than ATM. The distribution of estimated diffuse neutrino
flux from these young pulsars is narrower than AGNs and GRBs; the
flux from AGNs and GRBs peaks at about 20PeV and 1PeV
respectively(\citealt{He+etal+2012}). The neutrino production in
young pulsars is constrained by the accelerating efficiency of the
protons, while the AGNs are more powerful accelerators than pulsars.
Maybe the most violent processes in the universe such as AGNs or
GRBs can contribute in this energy range. The predicted diffuse
neutrino from pulsars in Galaxy(\citealt{Jiang+etal+2007}) is easier
to be detected, because the flux is above the sensitivity threshold
of IceCube. Why they obtained higher flux than us,because the polar
cap area we calculated is different.

\begin{figure}[h!!!]
   \centering
   \includegraphics[width=14cm, angle=0]{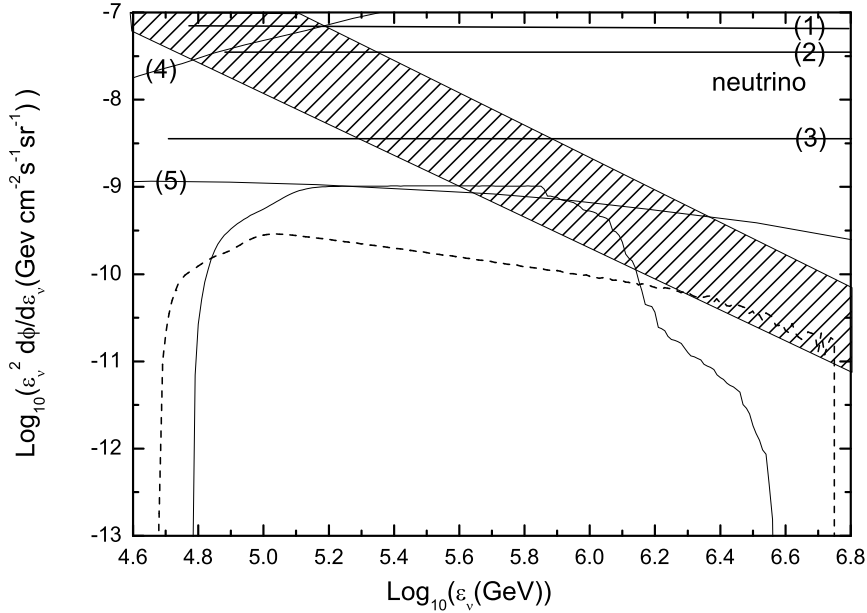}

   \begin{minipage}[]{120mm}

   \caption{The estimated diffuse neutrinos flux from young pulsars in our
Galaxy. Solid and Dot line correspond to the fluxes computed for the
cases of linear and quadratic proton acceleration. The solid lines
labeled (1),(2),and(3)represent the sensitivities of
AMANDA-II,ANTARES, and IceCube, respectively.The lines labeled
(4),(5)represent the predicted fluxes from
AGNs(\citealt{Stecker+etal+2005})and GRBs(\citealt{Liu+Wang+2013})
The region labeled "ATM" is the background atmospheric neutrino
flux(\citealt{Ribordy+etal+2006}).}
\end{minipage}
   \label{Fig5}
   \end{figure}

The results of diffuse gamma-rays flux are shown in fig.6,Solid and
Dashed line correspond to the fluxes computed for the cases of
linear and quadratic proton acceleration.

   \begin{figure}[h!!!]
   \centering
   \includegraphics[width=14.0cm, angle=0]{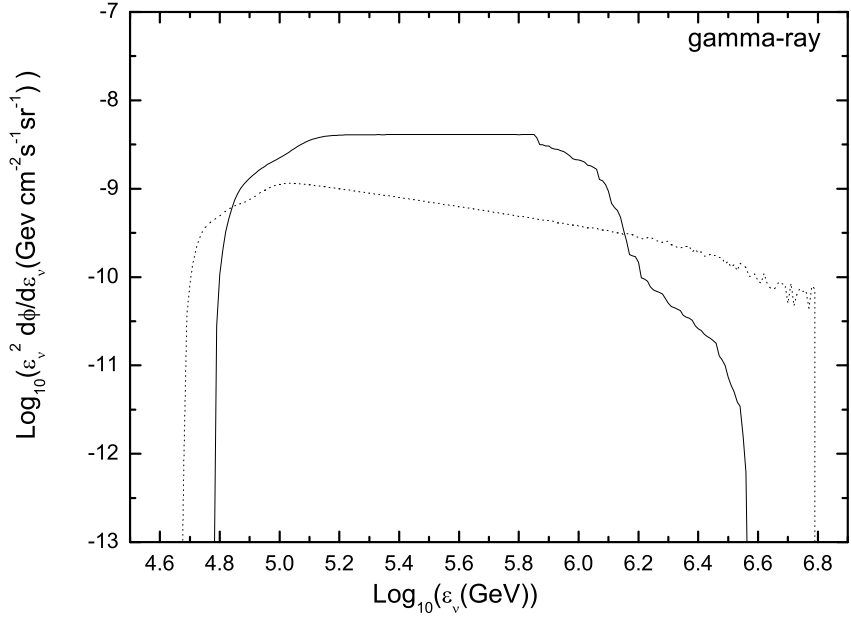}

   \begin{minipage}[]{120mm}

   \caption{ The estimated diffuse gamma ray flux from young pulsars in our
Galaxy. Solid and Dashed line correspond to the fluxes computed for
the cases of linear and quadratic proton acceleration. }
  \end{minipage}
   \label{Fig6}
   \end{figure}

\section{Discussion and Conclusion}

To summarize, based on the LB model and Bhadra(2009), we calculated
the neutrinos and gamma-ray spectrum of some pulsars with
appropriate polar cap area. we use MC method simulate pulsar sample
with ages less than $10^6yr$ and
 $\mu\cdot\Omega<0$ in our galaxy. Compared with
the ATNF catalogue pulsars, our simulation is succeed. According to
the sample, we also estimated the diffuse neutrinos and gamma-rays
in our galaxy. Our results show that the diffuse flux distributing
in the energy range from $\sim50TeV$ to $\sim3PeV$ is lower about 3
magnitude than Jiang(2007)only due to different methods on dealing
polar cap area.

These results show that  the signals of neutrino flux from young
pulsars are not easier to be detected, because the energy range
about $\sim50TeV$ to $\sim3PeV$ is lower than the sensitivity
threshold of IceCube and the atmospheric background neutrino flux.
Presently, any significant statistically signals excess have not
been detected from the direction of any pulsar by the AMANDA-II
telescope(\citealt{Ackermann+etal+2005,Ackermann+etal+2008};
\citealt{Ahrens+etal+2004}). So proper consideration of the effect
of polar cap geometry in flux calculation is important and pulsars
are unlikely to be strong source of TeV neutrinos. Compared with the
TeV diffuse neutrinos flux emitted from
AGNs(\citealt{Stecker+etal+2005}) and GRBs(\citealt{Liu+Wang+2013}),
AGNs maybe be the primary TeV neutrinos source.

 The energy data on the diffuse gamma-rays fluxes from Galactic plane around
$\sim50TeV$ to $\sim3PeV$ have not been found now. So we cannot give
the percentage of gamma-rays produced through $\Delta$-resonance
among the whole diffuse gamma-rays in our galaxy. We expect
experiments like the proposed High Altitude Water Cherenkov (HAWC)
detector will  be constantly used to survey large regions of the
sky, in particular the Galactic plane, at gamma-ray energies up to
100TeV with 10 to 15 times the sensitivity of Milagro.  If we have
the energy data on the diffuse gamma-rays fluxes from the Galactic
plane around $\sim50TeV$ to $\sim3PeV$. We can estimate the
percentage of pulsar's contribution on this energy range.

 This work have some uncertainties, such as pulsar's birth rate
and how many pulsars satisfy $\mu\cdot\Omega<0$.
 Estimating thermal photons from the neutron
star surface is not well determined and  the physical processes in
the magnetosphere are still not clear, resulting in some
uncertainties in the LB model as mentioned by
(\citealt{Link+Burgio+2005,Link+Burgio+2006}). Therefore, in order
to improve the prediction of the diffuse muon neutrino and gamma-ray
flux from young pulsars, it is crucial to understand more about the
neutrino flux and spectra emitted by a Single pulsar.
\begin{acknowledgements}
The authors are indebted to Professor Ze-Jun Jiang for his
constructive ideas and helpful suggestions on the manuscript. We
thank the referee for helpful comments and suggestions. This work is
partially supported by Science Research Foundation of Yunnan
Education Department of China(2012Y316)and Yunnan Province Grant No.
2010CD112.

\end{acknowledgements}

\bibliographystyle{raa}
\bibliography{bibtex}

\begin{thebibliography}{30}
\providecommand{\natexlab}[1]{#1}
\providecommand{\selectlanguage}[1]{\relax}

\bibitem[{Abbasi et~al.(2012)Abbasi, Abdou, Abu-Zayyad
  et~al.}]{Abbasi+etal+2012}
Abbasi, R., Abdou, Y., Abu-Zayyad, T., et~al. 2012, The Astrophysical Journal,
  745, 45

\bibitem[{{Ackermann} et~al.(2008){Ackermann}, {Adams}, {Ahrens}, {Andeen}, \&
  {Auffenberg}}]{Ackermann+etal+2008}
{Ackermann}, M., {Adams}, J., {Ahrens}, J., {Andeen}, K., \& {Auffenberg}, J.
  2008, \apj, 675, 1014

\bibitem[{Ackermann et~al.(2005)Ackermann, Ahrens, Bai
  et~al.}]{Ackermann+etal+2005}
Ackermann, M., Ahrens, J., Bai, X., et~al. 2005, Phys. Rev. D, 71, 077102

\bibitem[{{Aharonian} et~al.(2006){Aharonian}, { Akhperjanian}, {Bazer-Bachi},
  \& {Beilicke}}]{Aharonian+etal+2006a}
{Aharonian}, F., { Akhperjanian}, A.~G., {Bazer-Bachi}, A.~R., \& {Beilicke},
  M. 2006, \aap, 457, 899

\bibitem[{Ahrens et~al.(2004)Ahrens, Bai, Barwick et~al.}]{Ahrens+etal+2004}
Ahrens, J., Bai, X., Barwick, S.~W., et~al. 2004, Phys. Rev. Lett., 92, 071102

\bibitem[{{Arons} \& {Scharlemann}(1979)}]{Arons+etal+1979}
{Arons}, J., \& {Scharlemann}, E.~T. 1979, \apj, 231, 854

\bibitem[{{Bednarek} et~al.(2005){Bednarek}, { Burgio}, \&
  {Montaruli}}]{Bednarek+etal+2005}
{Bednarek}, W., { Burgio}, G.~F., \& {Montaruli}, T. 2005, \apj, 49, 1

\bibitem[{{beskin} et~al.(1993){beskin}, {Gurevich}, \&
  {Istomin}}]{beskin+etal+1993}
{beskin}, V.~s., {Gurevich}, J.~C., \& {Istomin}, N. 1993, {Physics of the
  Pulsar Magnetosphere} (Cambridge Univ), 117

\bibitem[{{Bhadra} \& {Dey}(2009)}]{Bhadra+Dey+2009}
{Bhadra}, A., \& {Dey}, R.~K. 2009, \mnras, 395, 1371

\bibitem[{{Cheng} et~al.(1986){Cheng}, {Ho}, \& {Ruderman}}]{Cheng+etal+1986}
{Cheng}, K.~S., {Ho}, C., \& {Ruderman}, M. 1986, \apj, 300, 500

\bibitem[{{Cheng} \& {Zhang}(1998)}]{Cheng+etal+1998}
{Cheng}, K.~S., \& {Zhang}, L. 1998, \apj, 498, 327

\bibitem[{{Daugherty} \& {Harding}(1996)}]{Daugherty+etal+1996}
{Daugherty}, J.~K., \& {Harding}, A.~K. 1996, \apj, 458, 278

\bibitem[{{Gaisser} et~al.(1995){Gaisser}, {Halzen}, \&
  {Stanev}}]{Gaisser+etal+1995}
{Gaisser}, T.~K., {Halzen}, F., \& {Stanev}, T. 1995, Phys. Rep, 258, 173

\bibitem[{{Goldreich} \& {Julian}(1969)}]{Goldreich+etal+1969}
{Goldreich}, P., \& {Julian}, W.~H. 1969, \apj, 157, 869

\bibitem[{{Gonthier } et~al.(2002){Gonthier }, {Ouellette}, {Berrier},
  {O'Brien}, \& {Harding}}]{Gonthier+etal+2002}
{Gonthier }, P.~L., {Ouellette}, M.~S., {Berrier}, J., {O'Brien}, S., \&
  {Harding}, A.~K. 2002, \apj, 565, 482

\bibitem[{He et~al.(2012)He, Liu, Wang et~al.}]{He+etal+2012}
He, H.-N., Liu, R.-Y., Wang, X.-Y., et~al. 2012, \apj, 752, 29

\bibitem[{{Hill}(2006)}]{Hill+2006}
{Hill}, G.~C. 2006, astro.ph, 773

\bibitem[{{Jiang} et~al.(2007){Jiang}, {Cheng}, \& {Zhang}}]{Jiang+etal+2007}
{Jiang}, Z.~J., {Cheng}, L.~B., \& {Zhang}, L. 2007, \apj, 667, 1059

\bibitem[{{Jiang} \& {Zhang}(2006)}]{Jiang+etal+2006}
{Jiang}, Z.~J., \& {Zhang}, L. 2006, \apj, 643, 1130

\bibitem[{Kistler \& Beacom(2006)}]{Kistler+etal+2007}
Kistler, M.~D., \& Beacom, J.~F. 2006, Phys. Rev. D, 74, 063007

\bibitem[{{Link} \& {Burgio}(2005)}]{Link+Burgio+2005}
{Link}, B., \& {Burgio}, F. 2005, Phys. Rev. Lett., 94, 181101

\bibitem[{{Link} \& {Burgio}(2006)}]{Link+Burgio+2006}
{Link}, B., \& {Burgio}, F. 2006, \mnras, 371, 375

\bibitem[{Liu \& Wang(2013)}]{Liu+Wang+2013}
Liu, R.-Y., \& Wang, X.-Y. 2013, \apj, 766, 73

\bibitem[{{Montaruli}(2005)}]{Montaruli+2005}
{Montaruli}, T. 2005, Acta Phys.Polon.B, 36, 509

\bibitem[{{Ribordy}(2006)}]{Ribordy+etal+2006}
{Ribordy}, B. 2006, Phy.Atom.Nucl, 69, 1899

\bibitem[{{Ruderman} \& {Sutherland}(1975)}]{Ruderman+etal+1975}
{Ruderman}, B.~A., \& {Sutherland}, P.~G. 1975, \apj, 196, 51

\bibitem[{Stecker(2005)}]{Stecker+etal+2005}
Stecker, F.~W. 2005, Phys. Rev. D, 72, 107301

\bibitem[{{Sturner} \& {Dermer}(1996)}]{Sturner+etal+1996}
{Sturner}, S.~J., \& {Dermer}, C.~D. 1996, \apj, 461, 872

\bibitem[{{Zhang} \& {Harding}(2000)}]{Zhang+etal+2000}
{Zhang}, B., \& {Harding}, A.~K. 2000, \apj, 532, 1150

\bibitem[{{Zhang} \& {cheng}(2004)}]{Zhang+etal+2004}
{Zhang}, L., \& {cheng}, K.~S. 2004, \apj, 604, 317

\end{thebibliography}
\end{document}